\documentclass[a4paper,11pt]{article}
\usepackage{amsmath}

\usepackage{jheppub} 

\usepackage[T1]{fontenc} 

\title{\boldmath Non-Degenerate Squarks from Flavored Gauge Mediation}

\author[a]{Iftah Galon,}
\author[b,c]{Gilad Perez,}
\author[a]{and Yael Shadmi}

\affiliation[a]{Physics Department, Technion---Israel Institute of
Technology,\\ Haifa 32000, Israel}
\affiliation[b]{CERN Theory Division, CH-1211, Geneva 23, Switzerland}
\affiliation[c]{Department of Particle Physics and Astrophysics, \\ 
+Weizmann Institute of Science, Rehovot 76100, Israel}

\emailAdd{iftah@tx.technion.ac.il }
\emailAdd{gilad.perez@cern.ch}
\emailAdd{yshadmi@physics.technion.ac.il}

\abstract{
We study the squark spectra of Flavored Gauge Mediation
Models, in which messenger-matter superpotential couplings
generate new, generation-dependent contributions to the squark masses.
The new couplings are controlled by the same flavor symmetry
that explains the fermion masses, leading to excellent alignment
of the quark and squark mass matrices. This allows for large 
squark mass splittings consistent with all flavor bounds.
In particular, second-generation squarks are often significantly
lighter than the first-generation squarks. 
As squark production 
at the LHC is dominated by the up- and down-squarks 
and the efficiencies for squark searches increase with their masses, 
the charm and/or strange squark
masses can be well below the current LHC bounds.
At the same time, even with a single set of messengers, the models 
can generate large stop mixings which result in large loop
contributions to the Higgs mass. 
}

\newcommand{\PRE}[1]{}       
\usepackage{hyperref,rotating}      
\usepackage{subfigure}

\renewcommand{\eqref}[1]{eqn.~(\ref{#1})}
\newcommand{\eqsref}[2]{eqs.~(\ref{#1}) and (\ref{#2})}

\newcommand{\secref}[1]{Sec.~\ref{sec:#1}}

\newcommand{\appref}[1]{Appendix~\ref{sec:#1}}
\newcommand{\figref}[1]{Fig.~\ref{fig:#1}}

\newcommand{\tableref}[1]{Table~\ref{#1}}

\newcommand{\bea}{\begin{eqnarray}}
\newcommand{\eea}{\end{eqnarray}}
\newcommand{\beq}{\begin{equation}}
\newcommand{\eeq}{\end{equation}}
\newcommand{\beqa}{\begin{eqnarray}}
\newcommand{\eeqa}{\end{eqnarray}}
\newcommand{\nn}{\nonumber}

\def\lsim{\mathrel{\rlap{\lower4pt\hbox{\hskip1pt$\sim$}}   
\raise1pt\hbox{$<$}}}                
\def\gsim{\mathrel{\rlap{\lower4pt\hbox{\hskip1pt$\sim$}}    
\raise1pt\hbox{$>$}}}                

\begin{document}


\maketitle

\section{Introduction}
\label{sec:introduction}
As ATLAS and CMS are excluding large portions of the parameter
space of supersymmetric extensions of the standard model, it is
important to ensure that superpartners are not escaping detection
simply because current searches are optimized for specific
spectra. One common assumption
 is that first- and second-generation squarks are 
degenerate.
There are various viable schemes however for the mediation of 
supersymmetry breaking
in which this assumption does not hold, leading do different 
and distinct collider signatures~\cite{
Feng:2007ke,
Kribs:2007ac,
Nomura:2008pt,
Nomura:2008gg,
Feng:2009bd,
Shadmi:2011hs,
Galon:2011wh}.
A particularly intriguing possibility is that
second-generation squarks are
significantly lighter than 
the current ${\cal O} (1.5\,\rm TeV)$ 
bounds~\cite{
CMS-PAS-SUS-12-018,
ATLAS-CONF-2013-026,
Chatrchyan:2013lya,
Aad:2013oua,
ATLAS-CONF-2013-047,
ATLAS-CONF-2013-054,
ATLAS-CONF-2013-061,
ATLAS-CONF-2013-062}
which are mainly sensitive to the ``valence" squark masses and 
are optimized to look for heavy squarks~\cite{Mahbubani:2012qq}. 
In fact recasting existing analyses it was found in~\cite{Mahbubani:2012qq} 
that a single weak singlet of the second generation such as the charm or 
strange squark as light as ${\cal O} (400\,\rm GeV)$ is consistent with
current direct searches. The mixing of such a state with the stops is 
also unconstrained at present and would weaken the
bound on their masses~\cite{Blanke:2013uia}.
 
Non-degenerate TeV-scale first- and second-generation squarks
can be consistent with bounds on 
flavor-violating processes in alignment models~\cite{Nir:1993mx,Leurer:1993gy}.
In these models, both the fermion and sfermion mass matrices are
controlled by a broken flavor symmetry, so that they are approximately 
diagonal in the supersymmetry interaction basis.
Specifically,
the mass-squared matrices for the SU(2)-singlet  up (down) squarks 
must be aligned with the up (down) singlet quark Yukawa matrices, 
while the doublet squarks mass-squared matrix needs to be aligned with 
the down-quark Yukawa matrix. It was pointed out recently that, 
despite the considerable progress in constraining 
CP violation in $D-\bar D$ mixing (see~\cite{Aaij:2012nva}),
alignment models can be consistent with all 
flavor constraints~\cite{Blum:2009sk,Gedalia:2012pi, Kadosh-Paradisi-Perez}.
However, 
known  examples of alignment 
are typically high-scale models,
with supersymmetry-breaking mediated to the Minimal 
Supersymmetric Standard Model (MSSM) at scales
close to the GUT scale. The gluino mass then induces a large universal
contribution to the squark masses through the RGE evolution to the weak scale. 
Since this contribution makes up  80-85\% to the squark masses,
these models can only lead to modest non-degeneracies
in the squark spectrum.

An alternative scenario which allows for sfermion non-degeneracy
with flavor violation suppressed by alignment, is Flavored Gauge 
Mediation (FGM)~\cite{Shadmi:2011hs}.
FGM models are simple variations of 
Minimal Gauge Mediation models~\cite{Dine:1994vc,Dine:1995ag},
with superpotential couplings of the messengers  to the matter fields.
These couplings generate new, generation-dependent contributions to the 
soft masses, which are nonetheless consistent with flavor bounds,
if the resulting contributions are aligned with the fermion matrices.
This possibility can be realized in the context of flavor symmetries
as in~\cite{Nir:1993mx}. However,  while in the models of~\cite{Nir:1993mx},
the flavor symmetry controls the soft-supersymmetry breaking terms,
FGM models exhibit {\sl supersymmetric alignment},
with the flavor symmetry  controlling the new superpotential couplings.
Thus, alignment can be achieved in these models even for 
a low supersymmetry breaking scale, for which RGE effects are
small.

Gauge Mediated Supersymmetry Breaking (GMSB) models with 
messenger-matter couplings were originally studied 
in~\cite{Dvali:1996cu,Dine:1996xk,Giudice:1997ni,Chacko:2001km, Joaquim:2006uz, Joaquim:2006mn, Brignole:2010nh}.
Unlike pure GMSB models~\cite{earlygmsb},
these models have A-terms at the messenger scale.
In addition, the new contributions to the stop masses are negative
in large parts of the parameter space.
As a result, stop mixing can be significant in these models,
which allows for a heavy Higgs with relatively light 
superpartners~\cite{Evans:2011bea,
Evans:2011uq,
Evans:2012hg,
Kang:2012ra,
Evans:2012uf,
Craig:2012xp,
Albaid:2012qk,
Abdullah:2012tq,
Craig:2013wga,
Evans:2013kxa,
Calibbi:2013mka}.

In~\secref{FGM} we describe the structure of FGM models,
and explain the qualitative behavior of the new contributions
to the squark masses-squared. These contributions start at one-loop, 
but the one-loop results are sub-leading in the supersymmetry
breaking. Thus these one-loop contributions dominate at low messenger scales,
where the universal RGE contribution is smallest.
Since the one-loop contributions are always negative, the masses
are always driven lower in this part of the parameter space.
At higher messenger scales, roughly above 10$^7$~GeV, the two-loop
contributions dominate. These are negative for small messenger-matter couplings,
and change sign for couplings of order unity.
We also collect in this section the relevant flavor bounds,
and discuss their implications for our models.

In~\secref{non_deg} we give several examples, for a single set of messengers 
with up-type messenger couplings. The models are then characterized 
by the charge assignments of the messengers under the flavor symmetry.
These lead to diverse and interesting spectra, including examples in
which the charm and strange squarks are significantly lighter than 
the GMSB predictions,
or conversely, some of the up and down quarks are heavier than
the GMSB predictions. 

Interestingly, even with a single set of messengers, the new couplings
can generate large contributions to the stop masses and A-terms 
together with large and negative contributions to the charm squark.
Thus, these models can accommodate a 126~GeV Higgs mass with a non-degenerate
squark spectrum.

The relevant expressions for the soft terms are collected 
in~\appref{soft_terms}.

\section{Setup and general considerations}
\label{sec:FGM}
In Flavored Gauge Mediation Models~\cite{Shadmi:2011hs}, 
the messenger fields have superpotential
couplings to the MSSM matter fields, leading to new, generation dependent
contributions to the soft terms. 
The models below are based on minimal GMSB 
models~\cite{Dine:1994vc,Dine:1995ag}, with a 
single supersymmetry-breaking spurion
\begin{equation}
\langle X\rangle = M+F\theta^2\,,
\label{Breaking}
\end{equation}
and $N_5$ pairs 
of messengers  transforming as 5$+\bar5$ of SU(5).
We use capital letters to denote the messenger fields, with
$5=T+D$ and $\bar 5=\bar T+\bar D$,  
where $T$ ($\bar T$) and $D$ ($\bar D$) are
fundamentals (anti-fundamentals) of $SU(3)$ and $SU(2)$ respectively. 

We will only consider up-type messenger couplings, with the superpotential
given by,
\beq
\label{eq:superpot}
W = X \left(\bar{T}_I T_I + \bar{D}_I  D_I\right)+  
H_U\, q\, Y_U \,  u^c +  H_D\, q\, Y_D\, d^c +  H_D\, l\, Y_L\, e^c +
 \bar D q \, y_U\,  u^c  \,.
\eeq
Here $q$, $u^c$, $d^c$, $l$, $e^c$ are the MSSM matter superfields,
$Y_U$, $Y_D$, $Y_L$ are the MSSM Yukawa matrices,
$y_U$ are the up-type messenger-matter Yukawa matrices,
$I=1,\ldots,N_5$ runs over the messenger pairs, and
$\bar D \equiv \bar D_1$ denotes the messenger field 
with superpotential couplings to matter.

Note that we have chosen the $D$ and $\bar D$ messengers to have the same
R-parity as the Higgses. There are different choices of symmetries which 
result in the structure of~\eqref{eq:superpot}, and forbid generic Higgs 
couplings to the messengers. 
One example is the standard U(1) Peccei-Quinn symmetry that forbids
the mu-term, with for example, the Higgses carrying charge 2 and the matter
fields carrying charge $-1$ each. Since we want $\bar D$ couplings
to the matter fields, $\bar D$ must also carry charge 2, with $D_1$
carrying charge $-2$. Just as in minimal GMSB models, this symmetry
still allows for explicit messenger mass terms. To forbid these as
well, one can employ another $U(1)$ or $Z_3$ symmetry, with the messengers
transforming as in~\tableref{tab:new_sym_charges}.
\begin{table}[h]
\centering
\renewcommand{\arraystretch}{1.25}
\begin{tabular}{|c||c|c|c|c|c|c|}
\hline
Superfield & $X$ & $D_{I\neq2}$ & $D_2$ 
& $\bar D_1$ & $\bar D_{I\neq1}$ & $T_I, \bar{T}_I$ \\
\hline
$Z_3$ & $1$ & $-1$ & $0$ & $0$ & $-1$ & $1$ \\
\hline
\end{tabular}
\caption{$Z_3$ charge assignments.}
\label{tab:new_sym_charges}
\end{table}
In the following we will additionally impose a flavor symmetry
with different $\bar D$ and $H_U$ charges. 
While these charges typically forbid the superpotential
coupling $X D H_U$,
there is always $\bar D-H_U$ mixing
from the K\"ahler potential, 
\beq
K \supset \epsilon \,\bar D^\dagger H_U + {\rm h.c.}
\eeq  
where $\epsilon$ is given by a power of the 
spurion field
which breaks the flavor symmetry. 
We can always redefine the fields to obtain a canonical K\"ahler potential,
and subsequently identify the combination that couples to $X$ as the messenger
$\bar D$, and the orthogonal combination as $H_U$\footnote{This last step
involves a 
rotation of the canonically normalized fields.}.
As a result,
the coupling $y_U$ is modified as,
\beq
y_U\rightarrow y_U +\epsilon Y_U\,.
\eeq
This modification will have very small effects in the  models 
below\footnote{Note that the shift in $y_U$ is proportional
to the matrix $Y_U$ and is therefore an MFV effect.
}.

The messenger-matter couplings $y_U$ generate new 
contributions to the soft masses.
These contributions were calculated for the 3-generation case
in~\cite{Abdullah:2012tq} (see also~\cite{Evans:2013kxa,
Calibbi:2013mka})
and are collected 
in~\appref{soft_terms}. 
The new contributions have a few noteworthy features.
First, the new couplings generate A-terms at one-loop.
If these have large 33 entries, the MSSM lower bound on the stop
masses required for a 126~GeV Higgs can be significantly relaxed. 
These $A$ terms will
have very small entries for the first and second generations in 
all of our models.
Second, the scalar masses-squared receive 
2-loop contributions from the new couplings, which, just like the
pure GMSB contributions, appear at leading order
in the supersymmetry breaking, ${\cal O}(F^2/M^2)$. 
These involve $y^4$ terms, mixed
gauge-$y^2$ terms 
and mixed $y^2-Y^2$ terms.
The latter can have
either sign, and their effects on
the first and second generation squarks are 
usually negligible. 
On the other hand, when the matrix $y_U$ has a single dominant
entry, as will be the case in all our examples,
the former terms have definite signs: the $y^4$ contributions
are positive and the gauge-$y^2$ contributions are negative.
In addition, 
there are  one-loop contributions at ${\cal O}(F^4/M^6)$,
which are always negative.
Since the ratio of the one-loop to the two-loop contribution
scales as $F^2/M^4$, and $F/M$ gives the overall scale of the
soft masses, the one-loop contributions are only important
at low messenger scales.

\begin{figure}[t] 
     \subfigure[] {
                \centering
                \includegraphics[width=0.5\textwidth]{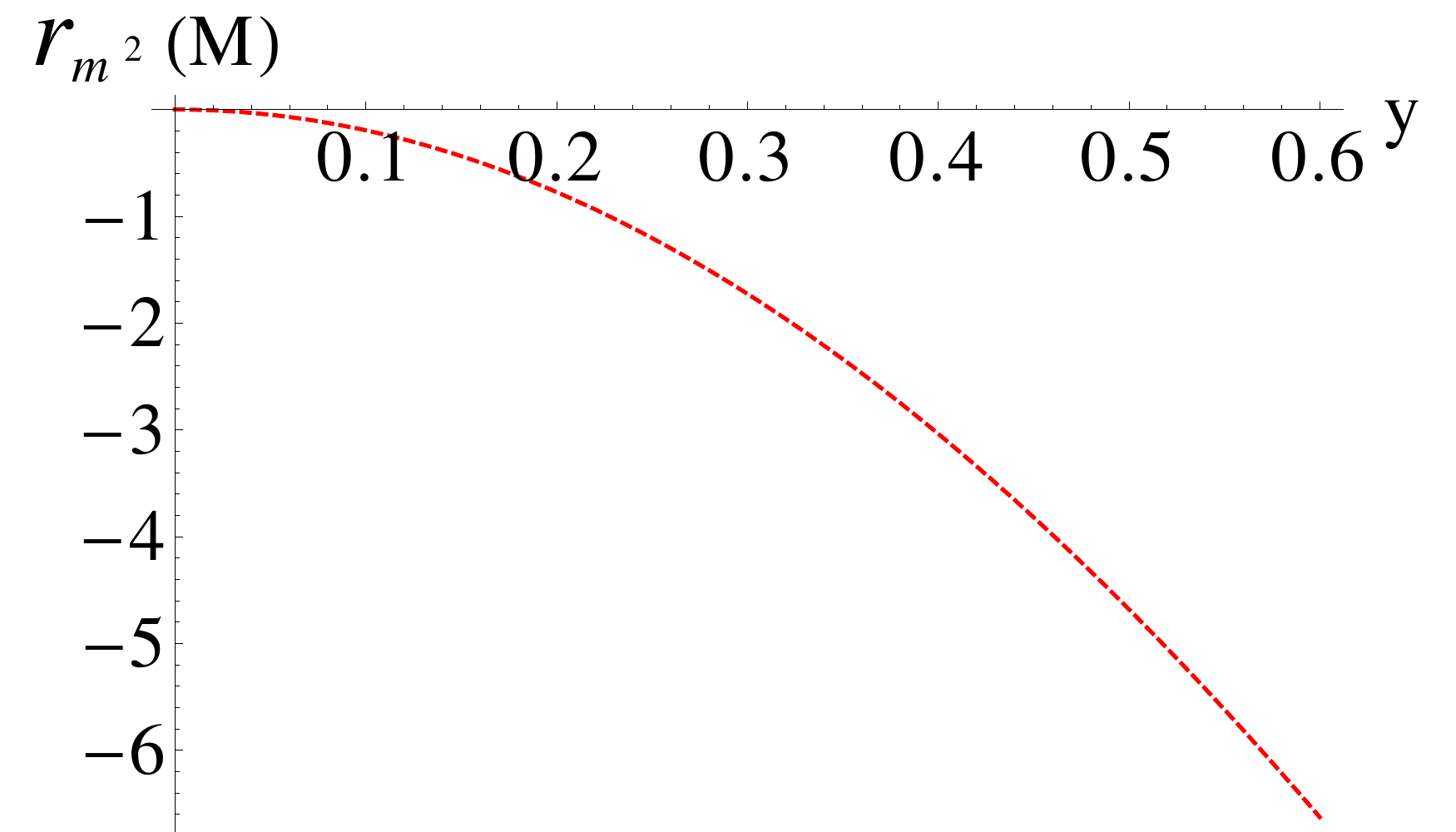}
		\label{fig:relb}
	}
 \subfigure[]{
                \centering
                \includegraphics[width=0.5\textwidth]{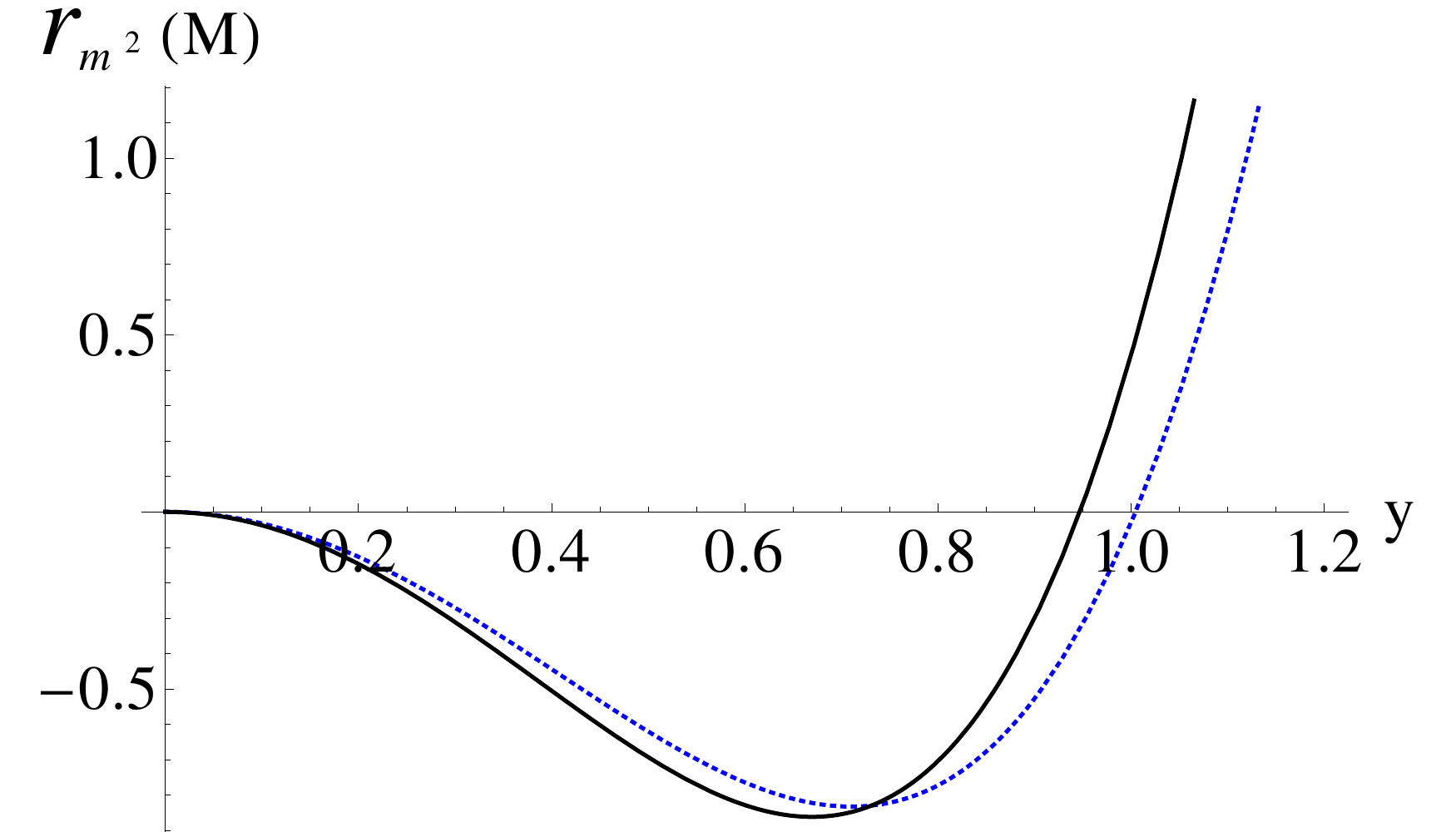}
		\label{fig:rela}
	}
\caption{The ratio of the new contribution to the GMSB contribution, 
$r_{m^2}$, 
with $F/M=8 \cdot10^4$~GeV  and $N_5=1$, 
for three different values of the messenger scale $M$.
In 1a. we show the 1-loop contribution for
$M=10^5$~GeV. In 1b. we show the 2-loop contribution for 
$M=10^6$~GeV (dotted, blue), and
$M=10^8$~GeV (solid,black).}
\label{fig:rel}
\end{figure}

It is instructive to examine the new contributions in the simple case
that only the 11 or 22 entry of $y_u$ is non-zero. This will indeed be the case
in most of our examples below\footnote{In fact, because the first-
and second-generation Yukawas are negligible, the relevant quantities
are $y_U y_U^\dagger$ and  $y_U^\dagger y_U$ (see~\eqref{sqmasses}), 
and all we need to assume here
is that these matrices have a single entry on the diagonal.}.
Taking for example
\beq\label{eq:singley}
{y_u}_{ij}=y\, \delta_{i2}\, \delta_{j2} \,,
\eeq
with all other entries zero or negligible,
the only change in the squark mass matrices is
\beq
\label{eq:split0}
{(\delta m_q^2)}_{22}=  \frac12{(\delta m_u^2)}_{22}\equiv \delta m^2 
\,,
\eeq
with,
to leading order in $F/M^2$, 
\beq\label{eq:split1}
\delta m^2 \sim
-\frac{1}{(4\pi)^2}
\frac{1}{6}
\vert y\vert^2
 \frac{F^4}{M^6}+
\frac{1}{(4\pi)^4}
\left(6\vert y\vert^2-G_y\right)\vert y\vert^2
\frac{F^2}{M^2}
\,,
\eeq
where
\beq
G_y \equiv
\frac{16}{3}g_3^2+3g_2^2+\frac{13}{15}g_1^2 \,.
\eeq
In~\figref{rel}, we plot the relative change in the squark mass squared
at the messenger scale, 
\beq\label{splitting0}
r_{m^2}(M)\equiv \frac{\delta{m^2}}{{m_{q,GMSB}^2}} 
\,,
\eeq
where ${m_{q,GMSB}^2}$ is the pure GMSB contribution
at different messenger scales.
For low messenger scales~(\figref{relb}), the negative one-loop
contribution dominates, and the shift in the squark mass squared
is sizeable even for low values of $y$. Thus the most interesting
range of $y$'s at these low scales is around 0.2 or so, which is
naturally obtained with a single Cabibbo suppression.
For larger messenger scales~(\figref{rela}), 
the new contribution changes sign
near $y=1$, so that there are two qualitatively different regions.
For small $y$'s, say 0.2 to 0.8,  the new contribution is negative
and the relative mass splitting $r_{m^2}$ varies
roughly between $-0.5$ to $-1$. 
Since our flavor spurion will be roughly 0.2, there is of course 
some ambiguity 
as to whether values of $y$ near 0.8 should be thought of
as one or zero powers of the spurion.   
For $y>1$, the new contribution grows
very fast, but such high values of the coupling are undesirable
anyway from the point of view of perturbativity.

For an arbitrary coupling matrix $y_U$, the new contributions
would lead to severe SUSY flavor and  CP problems.
However, as stressed in~\cite{Shadmi:2011hs},
any flavor theory that controls the Standard Model (SM) Yukawas will necessarily
generate a non-trivial structure of the new couplings.
In particular, if fermion masses are explained by a flavor symmetry,
the new couplings will be determined (up to order one coefficients)
once the flavor charges of the messengers are specified.
Since we are mainly interested in large effects in the first and second
generation squarks, our examples will be based on
different flavor charges for $\bar D$ and $H_U$, such that $y_U$ has large
entries in the first- and second-generation.

For convenience, we collect here the relevant bounds on flavor-violation 
involving the first two generations and $b\to s$ transitions
from~\cite{Isidori:2010kg}\footnote{The bounds quoted here are a bit 
stronger compared to~\cite{Isidori:2010kg} as we assume order one 
CP-violating phases, as is generically expected in our framework. 
Furthermore, the bounds related to 23 mixing are taken to be $\sim 3$ 
stronger compared to~\cite{Isidori:2010kg} due to the recent LHCb 
constraints on CP violation in $B_s$ mixing~\cite{Aaij:2013iua}.}.
Since we are interested in LHC phenomenology, it is useful to work
in the squark mass basis, and express the constrained quantities
in terms of the squark masses, $m_{q_i}$, and the quark-squark-gluino couplings
$K_{ij}$. The constrained quantities are then, 
\beq
(\delta^q_{ij})_{MM} = 
\frac{\Delta\tilde m^2_{q_j q_i}}{\tilde m_q^2} 
(K^q_M)_{ij} (K^q_M)^*_{jj},
\eeq
where 
$\Delta \tilde m^2_{q_j q_i} = m^2_{q_j} - m^2_{q_i}$, and
 $\tilde m^2_{q}$
is the average of relevant squark masses-squared of the same type.
The experimental constraints on the $\delta$'s as 
derived in~\cite{Isidori:2010kg} are summarized in~\tableref{tab:bounds}.
\begin{table}[h]
\centering
\renewcommand{\arraystretch}{1.25}
\begin{tabular}{|cc|ccc|}
\hline
$q$ & $ij$ & $| (\delta^q_{ij})_{MM}|$ & 
$\sqrt{{\rm Im}(\delta^q_{ij})_{MM}^2}$  & 
$ \sqrt{{\rm Im}((\delta^q_{ij})_{LL} (\delta^q_{ij})_{RR}})$ \\
\hline\hline
$d$ & 12& $0.07$ & $0.01$ & $0.0005$\\
$u$ & 12 & $0.1$& $0.05$ & $0.003$ \\
$d$ & 23 & $0.6$ & $0.2$ & $0.07$\\
\hline\hline
\end{tabular}
\caption{The upper bounds on $(\delta^q_{ij})_{M M}$,
taken from~\cite{Isidori:2010kg}, but assuming 
order-one phases, 
for $m_{\tilde q} = 1$~TeV and $m_{\tilde g} / m_{\tilde q} = 1$. 
}
\label{tab:bounds}
\end{table}
These bounds should be viewed as rough estimates only,
and will vary by order-1 numbers over the parameter space of our models.
In particular, while the values in~\tableref{tab:bounds}
were derived for common squark
and gluino masses, 
the gluino and different squark species have different masses in our models
with values given predominantly by the GMSB contributions.

The most severe constraint is on the product of the LL
and RR  1--2 mass splitting. Since the 
up 
LL and RR mass splittings
are typically of the same order of magnitude in our models, a strong suppression
of the 1--2 mixings is needed in order to allow for large mass differences.
This can be achieved with a U(1)$\times$U(1) flavor symmetry,
which can lead to a high level of down alignment.
Since the relative mass shift $r_{m^2}$ is suppressed
by $N_5$, and since the gluino mass generates a universal
contribution to the squark masses through the running,
and furthermore, the gluino to squark mass ratio scales as $\sqrt{N_5}$,
larger mass differences are obtained for $N_5=1$ and low messenger scales.

\section{Non-degenerate squarks}
\label{sec:non_deg}
\subsection{Flavor symmetry and fermion masses}
As explained above, in order to achieve sufficient suppression of
the 1--2 
mixing, we take the flavor symmetry to be
U(1)$\times$ U(1), 
with each U(1) broken by a spurion $\lambda\sim 0.2$ of charge $-1$.
Each Yukawa matrix element is then either suppressed 
by the appropriate power of  $\lambda$, or  
vanishes due to holomorphy.
Since the superpotential can only depend on the spurion $\lambda$,
and not $\lambda^\dagger$, superpotential terms of  negative total U(1)
charge cannot appear~\cite{Nir:1993mx}.

Following~\cite{Leurer:1993gy} we assign the $U(1)\times U(1)$ charges
\beqa
\begin{matrix}
Q_1(6,-3) & ,Q_2(2,0) & ,Q_3(0,0) \\
u^c_1(-6,9) & ,u^c_2(-2,3) & ,u^c_3(0,0) \\
d^c_1(-6,9) &, d^c_2(2,0) & ,d^c_3(2,0) \\
\end{matrix}
\eeqa
for the matter fields. The $H_U$, $H_D$ charges can always be 
chosen as zero~\cite{Leurer:1993gy}. 
The resulting SM Yukawas are
\beqa
Y_U \sim
\begin{pmatrix}
\lambda^6 & \lambda^4 & 0 \\
0 & \lambda^3 & \lambda^2 \\
0 & 0 & 1
\end{pmatrix}
\,,~
Y_D \sim
\begin{pmatrix}
\lambda^6 & 0 & 0 \\
0 & \lambda^4 & \lambda^4 \\
0 & \lambda^2 & \lambda^2
\end{pmatrix}
\,,
\eeqa
where here and throughout the paper, the different entries
of these matrices are known only up to order-one coefficients.
These lead to the fermion mixing matrices,
\beqa\label{eq:fermix}
V^u_L
&\sim&
\begin{pmatrix}
1 & \lambda & 0 \\
\lambda & 1 & \lambda^2 \\
0 & \lambda^2 & 1
\end{pmatrix}
\,,~
V^d_L
\sim
\begin{pmatrix}
1 & 0 & 0 \\
0 & 1 & \lambda^2 \\
0 & \lambda^2 & 1
\end{pmatrix}
\,,
\\
V^u_R
&\sim&
\begin{pmatrix}
1 & \lambda^4 & 0 \\
\lambda^4 & 1 & \lambda^5 \\
0 & \lambda^5 & 1
\end{pmatrix}
\,,~
V^d_R
\sim
\begin{pmatrix}
1 & 0 & 0 \\
0 & 1 &1 \\
0 & 1 & 1
\end{pmatrix}
\,.
\eeqa

The squark mass matrices are determined by the U(1)$\times$U(1)
flavor charges of the messengers. We choose these as
\beq
D(-n,m)\ \ \  \bar D(n,-m)
\eeq
where $n,m$ are integers. 
The different examples below correspond to different choices
of $n$ and $m$. For $m>0$, the resulting $y_U$ coupling 
will have large entries in the 1--2 block.
Furthermore, for 
$n>0$, the superpotential coupling
$X D H_U$ is forbidden. K\"ahler mixing of $H_U$ and $\bar D$ 
is suppressed by 
$\vert\lambda\vert^{n+m}$.

\subsection{Light charm- and strange-squarks}
Choosing $n=1$ and $m=-3$ we have
\beq
y_u
=
\begin{pmatrix}
\lambda^4 & 0 & 0 \\
0 & c_{22}\lambda & 0 \\
0 & 0 & 0
\end{pmatrix}
\,,
\eeq
where the zeros follow from holomorphy as described above,
and we explicitly displayed the order-one coefficient $c_{22}$ in 
the 22 entry.
This model is then precisely of the type~\eqref{eq:singley}.
The 22 entries of the LL and RR up squark mass-squared matrices
are modified as in~\eqsref{eq:split0}{eq:split1}, 
with $y=c_{22}\lambda$. All other entries of the up mass matrices,
as well as the down mass matrices, remain virtually unmodified.
For $c_{22}\sim1$, $\delta m^2$ can be large, so that the L and R charm squark
masses, as well as the L strange squark mass, are shifted to lower values.
While we loosely refer to the squark mass eigenstates as
``up'', ``charm'' etc., this is of course not quite accurate,
since the fermion mixing matrices of~\eqref{eq:fermix} introduce some
inter-generation mixings. Since the R down squark masses are unaltered,
the only potentially observable effect is the ${\cal O}(\lambda)$
12 mixing in the L up sector. Here and in the following we continue
to refer to the squark eigenstates by their dominant flavor component
whenever the remaining flavor components are indeed very small.

The A-terms (in the flavor basis) are given by
\beq
\delta A^*_u  \sim
-\frac{1}{16\pi^2}
\begin{pmatrix}
0 & \lambda^6 & 0\\
0 & \lambda^5 & \lambda^4 \\
0 & 0 & 0
\end{pmatrix}
\frac{F}{M}
,\qquad
\delta A^*_d \sim
-\frac{1}{16\pi^2}
\begin{pmatrix}
0 & 0 & 0 \\
0 & \lambda^6 & \lambda^6\\
0 & 0 & 0
\end{pmatrix}
\frac{F}{M}
\,,
\eeq
and can therefore be neglected\footnote{Furthermore,
because the matrix $y_U$ has a single entry, which enters in its
absolute value squared,
these A terms do not contain any new phases on top of the SM Yukawas.}.

While the $D$ and $\bar D$ charges forbid the superpotential term $X D H_U$,
they do allow $\bar D-H_U$ K\"ahler mixing proportional to $\lambda^4$.
As explained in~\secref{FGM}, the effect of this mixing is to modify
the new coupling as
\beq
y_U \rightarrow y_U + \lambda^4 Y_U\,.
\eeq
It is easy to see that this leads to negligible effects in all entries
of $y_u$ and the resulting soft terms.

Flavor constraints are satisfied in this model even for large mass splittings
thanks to the precise alignment of the down sector, which essentially 
eliminates any
CP violating contributions. 
The only non-negligible $\delta$'s are given by, 
\beqa
(\delta^u_{12})_{LL} &\sim& \lambda
\,r_{m^2}
,\qquad
(\delta^u_{23})_{LL} \sim \lambda^2 \,r_{m^2}
,\qquad
(\delta^d_{23})_{LL} \sim \lambda^2 \,r_{m^2}
,\qquad
\nn\\
(\delta^u_{12})_{RR} &\sim& \lambda^4 \,r_{m^2}
,\qquad
(\delta^u_{23})_{RR} \sim \lambda^5 \,r_{m^2}
,\qquad
\nn\\
(\delta^u_{12})_{LR} &\propto& \lambda^6
,\qquad
(\delta^u_{13})_{LR} \propto \lambda^5
,\qquad
(\delta^u_{23})_{LR} \propto \lambda^4
\,.
\eeqa
Here $r_{m^2}$ is the relative mass splitting as defined in~\eqref{eq:split1}
but evaluated at low energies. As the masses are evolved
down to the TeV scale, $r_{m^2}$ decreases, due to the universal
gaugino contribution.
The most dramatic effects are therefore obtained
for low messenger scales, and $N_5=1$.
Thus for example,
for $F/M=200$~TeV, $M=500$~TeV and $\tan\beta=5$, starting with 
$r_{m^2}(M)=-1$, all the squark masses are around 2~TeV,
the gluino mass is near 1.5~TeV, but the R-handed charm squark is
at 870~GeV. Taking instead $F/M=150$~TeV and $M=400$~TeV the gluino
is at 1.2~TeV, the R-handed charm squark at 670~GeV, while the 
remaining squarks are around 1.6~TeV.

To summarize, in large parts of the parameter space of this model
(given essentially by the coefficient $c_{22}$),
the two charm squarks and the left-handed strange 
squark are significantly lighter than the up and down squarks,
with flavor constraints satisfied by virtue of the precise down alignment.

\subsection{Heavy up- and down-squarks}
Taking $n=0$ and $m=6$ we again reproduce the coupling matrix $y_U$ of~\eqref{eq:singley}, but now with a nonzero 11 entry,
\beq
(y_U)_{ij} = y\, \delta_{1i}\, \delta_{1j}\,,
\eeq
where $y$ is order-1.
In the flavor basis,
the messenger contributions to the the soft masses are again as in~\eqref{eq:split0}
\beqa
(\delta \tilde m^2_{q})_{11} &=& \frac{1}{2}(\delta \tilde m^2_{u_R})_{11} =
\delta m^2
\,,
\\
\delta \tilde m^2_{d_R}  &\sim& 0
\,,
\eeqa
Since the valence squarks are shifted here, this model is mainly interesting
when $\delta m^2$ is positive. This implies that a.~the one-loop
contribution should be subdominant, which is the case for messenger scales
above roughly $10^7$~GeV, and b.~that $y$ should be close to 1
(see~\figref{rela}).
Of course, for high messenger scales, RGE effects
will significantly reduce the relative mass splittings.

The A-terms in this basis are given by
\beq
\delta A^*_u  \sim
-\frac{\vert y\vert^2}{16\pi^2}
\begin{pmatrix}
3\lambda^6 & \lambda^4 & 0\\
0 & 0 & 0 \\
0 & 0 & 0
\end{pmatrix}
\frac{F}{M}\,,
\qquad
\delta A^*_d  \sim
-\frac{\vert y\vert^2}{16\pi^2}
\begin{pmatrix}
\lambda^6 & 0 & 0 \\
0 & 0 & 0\\
0 & 0 & 0
\end{pmatrix}
\frac{F}{M}
\,.
\eeq
All entries in the A-terms are very small. 

Thus, 
 both the left-handed and right-handed squarks of the first generation
are split in mass from the second generation,
with the relative mass splitting 
$\delta m^2/m^2$
again as in~\eqref{eq:split1}.
For $y\gsim1$ the two up squarks and the left down squark are therefore heavier
than the remaining squarks and the relative mass splitting can be a large 
effect.
The strict bounds on flavor violation in the down sector are however satisfied
thanks to the excellent alignment of the down squark and quark mass matrices.
Since the squark mass matrix is diagonal in the flavor symmetry basis,
the mixings arise only from the fermion Yukawa matrices of~\eqref{eq:fermix}.
The only non-negligible $\delta$s are therefore
\beqa
(\delta^u_{12})_{LL} \sim \lambda  \,r_{m^2}
,\qquad
(\delta^u_{12})_{RR} \sim \lambda^4  \,r_{m^2}
,\qquad
(\delta^u_{12})_{LR} \propto \lambda^4  
\,,
\eeqa
and all of these are below the experimental bounds.
As in the previous model, the effects of 
the $\bar D - H_U$ kinetic mixing in the K\"ahler potential 
are negligible as they are proportional to $\lambda^6$.


\subsection{A light right-handed charm squark and a heavy Higgs}
Choosing $n=2$ and $m=3$
\beq
y_u
=
\begin{pmatrix}
\lambda^5 & 0 & 0 \\
0 & \lambda^2  & 0 \\
0& y & 0
\end{pmatrix}
\,,
\eeq
where $y={\cal O}(1)$, and as usual we do not display the  ${\cal O}(1)$
prefactors of the powers of $\lambda$ in the remaining terms.

Again we find that the right handed charm squark mass is shifted,
\beq
\delta (\tilde m^2_{u_R})_{22}  \sim
-\frac13\,\frac{1}{(4\pi)^2} \vert y\vert^2\frac{F^4}{M^6}
+\frac{1}{(4\pi)^4} 
2\left(
(6+|(Y_u)_{33}|^2) \vert y\vert^2-G_y
\right) \vert y\vert^2
\frac{F^2}{M^2}
\,,
\eeq
and this shift is negative in large regions of the parameter 
space\footnote{The right-handed down squark matrix 
has ${\cal O}(\lambda^4)$
entries in the 2--3 block, but these do not produce any significant effects.}.

At the same time however, because of the off-diagonal entry of $y_U$, 
there is also a large negative contribution 
to the right-handed stop mass,
\beq 
\delta (\tilde m^2_{u_R})_{33}  \sim
-2\vert y\vert^2|(Y_u)_{33}|^2
\frac{F^2}{M^2}\,.
\eeq
Note that this contribution is qualitatively
different than those commonly considered in studies focused on
raising the Higgs mass, which assume only a 33 entry in 
$y_U$~\cite{Evans:2011bea,
Evans:2011uq,
Evans:2012hg,
Kang:2012ra,
Evans:2012uf,
Abdullah:2012tq,
Craig:2013wga,
Evans:2013kxa,
Calibbi:2013mka}.
In particular, there is no positive $y_U^4$ contribution, or mixed gauge-Yukawa
contribution.

As for the doublet squarks, these involve the combination
\beq
y_U y_U^\dagger \sim 
\begin{pmatrix}
0 & 0 & 0 \\
0 & \lambda^4 & \lambda^2 \\
0 & \lambda^2 & \vert y\vert^2
\end{pmatrix}\,,
\eeq 
which, because the 3rd column of  $y_U$ vanishes,
has only one non-zero eigenvalue.
Thus only the masses of the third generation doublet 
squarks are affected,
with 
\beq
(\delta \tilde m^2)_{\tilde t_L}
=
(\delta \tilde m^2)_{\tilde b_L}
\sim
-\frac16\,\frac{1}{(4\pi)^2} \vert y\vert^2\frac{F^4}{M^6}+
\frac{1}{(4\pi)^4}
\left(
6\vert y\vert^2-G_y \right)\,\vert y\vert^2
\frac{F^2}{M^2} \,,
\eeq
with a 2--3 mixing of 
order $\lambda^2$.

The A-terms in the flavor basis are given by
\beqa
\delta A^*_u  \sim
-\frac{1}{16\pi^2}
\begin{pmatrix}
0 & \lambda^4 & 0  \\
0 & \lambda^3 & \lambda^2 \\
0 & \lambda^5 & \vert y\vert^2(Y_u)_{33}
\end{pmatrix}
\frac{F}{M}
,\qquad
\delta A^*_d  \sim
-\frac{1}{16\pi^2}
\begin{pmatrix}
0 & 0 & 0 \\
0 & \lambda^4 & \lambda^4 \\
0 & \lambda^2 & \lambda^2 
\end{pmatrix}
\frac{F}{M}
\,,
\eeqa
Thus, there is an order-1 stop A term. Coupled with the new negative 
contributions
to the stop masses, this leads to a large stop mixing which  enhances the
loop contributions to the Higgs mass.
Therefore, even with a single messenger pair,
the off-diagonal structure of the coupling $y_U$ results in  a
single light charm squark (the R-handed one), as well as a large 
Higgs mass.

The only non-negligible $\delta$s are
\beqa
(\delta^u_{12})_{LL} &\sim& \lambda^5
,\qquad
(\delta^u_{13})_{LL} \sim \lambda^3
,\qquad
(\delta^u_{23})_{LL} \sim \lambda^2
,\qquad
(\delta^d_{23})_{LL} \sim \lambda^2
,\qquad
(\delta^u_{12})_{RR} \sim \lambda^4
,\qquad
\nn\\
(\delta^u_{23})_{RR} &\sim& \lambda^5
,\qquad
(\delta^u_{12})_{LR} \propto \lambda^4
,\qquad
(\delta^u_{13})_{LR} \propto \lambda^3
,\qquad
(\delta^u_{23})_{LR} \propto \lambda^2
,\qquad
(\delta^d_{23})_{LR} \propto \lambda^5\,.
\nn\\
\eeqa

\section{Conclusions}
\label{sec:conclusions}
Flavored Gauge Mediation models provide a fully calculable
framework for generating the MSSM soft terms,
with the soft terms generated by the SM gauge interactions
and superpotential couplings of the messenger and matter superfields.
As we have seen, they allow for large mass splittings between
the squarks, with flavor constraints satisfied by alignment.
In particular, given that it is the superpotential couplings
that are controlled by the flavor symmetry,
this alignment can be achieved 
even at low messenger scales, reducing the large
logs and the resulting fine-tuning typical of  high-scale models.

The models lead to very interesting spectra from the point of view
of direct searches at the LHC. In particular, charm squarks can be significantly
lighter than the up and down squarks, and below current limits on the squark
mass. For a small number of messenger pairs, the charm
squark may be the only squark lighter than the gluino, so that
the gluino predominantly decays into on-shell charm squarks.
In addition, charm squarks generically have an ${\cal O}(\lambda)$ 
mixing with the up quark, which is typical of alignment 
models~\cite{Mahbubani:2012qq}. 
If charm tagging becomes possible in future
experiments, observing the different up and charm components of
the light charm squarks would be extremely interesting.

Finally, since the new couplings generate messenger-scale A-terms,
these models can lead both to large mass splittings
among the first and second generation squarks, {\sl and} to large contributions
to the Higgs mass, even with a single messenger pair.

\section*{Acknowledgments}
GP thanks  Gian Giudice, Paride Paradisi and Giovanni Villadoro 
for useful discussions.
The research of Y.~Shadmi and I.~Galon was supported by
the Israel Science Foundation (ISF) under grant No.~1367/11, 
and by the United States-Israel
Binational Science Foundation (BSF) under grant No.~2010221.
IG is also supported by the 2013 Gutwirth fellowship.
IG thanks the CERN TH division for their hospitality while this work was in
progress.
GP is supported by grants from GIF, IRG, ISF and Minerva.

\appendix
\section{Soft Terms}
\label{sec:soft_terms}
The pure GMSB contributions to the soft squared masses and
A-terms are given by~\cite{Dine:1994vc, Dine:1995ag}
\beqa
\tilde m^2_q 
&=&
\frac{1}{(4\pi)^4}
2N_5\left(\frac{4}{3}g_3^4 + \frac{3}{4}g_2^4 + \frac{1}{60}g_1^4\right)
\left|\frac{F}{M}\right|^2  
1_{3\times 3} \,,
\\
\tilde m^2_{u_R} 
&=&
\frac{1}{(4\pi)^4}
2N_5\left(\frac{4}{3}g_3^4 +  g_1^4\frac{4}{15}\right)
\left|\frac{F}{M}\right|^2  
1_{3\times 3} \,,
\\
\tilde m^2_{d_R}
&=&
\frac{1}{(4\pi)^4}
2N_5\left(\frac{4}{3}g_3^4  + \frac{1}{15}g_1^4\right)
\left|\frac{F}{M}\right|^2  
1_{3\times 3} \,,
\\
A_u &=& A_d = 0 \,.
\eeqa
where we have only shown the leading order terms in $F/M^2$.
As is well known~\cite{Martin:1996zb}, the higher order corrections
to these expressions are small.

In the presence of up type matter-messenger couplings~\cite{Abdullah:2012tq}
the squared-masses of squarks receive both 1-loop and 2-loop contributions
given by
\beqa\label{sqmasses}
\delta \tilde m^2_q 
&=&
-\frac{1}{(4\pi)^2}
\frac{1}{6}
\left(
y_u  y_u^{\dagger}
\right)
 \frac{F^4}{M^6} \, h(x)
\nn\\
&&
+\frac{1}{(4\pi)^4}
\Bigg\{
~~~~
\left(3Tr\left(y_u^{\dagger}y_u \right)
-\frac{16}{3}g_3^2 - 3g_2^2 -  \frac{13}{15}g_1^2\right) y_u y_u^{\dagger}
\nn \\
&&~~~~~~~~~~~~~~~~
+3y_u y_u^{\dagger}y_u y_u^{\dagger} 
+2y_u Y_u^{\dagger} Y_u y_u^{\dagger} 
-2Y_u y_u^{\dagger} y_u Y_u^{\dagger}
\nn \\
&&~~~~~~~~~~~~~~~~
\vphantom{\left(\frac{16}{3}\right)}
+y_u Y_u^{\dagger} Tr\left(3y_u^{\dagger} Y_u \right)
+Y_u y_u^{\dagger} Tr\left(3Y_u^{\dagger}y_u \right)
~~~~\Bigg\}
\left|\frac{F}{M}\right|^2  \,,
\\
\delta\tilde m^2_{u_R} 
&=&
-\frac{1}{(4\pi)^2}
\frac{1}{3}
\left(
y_u^\dagger  y_u
\right)
 \frac{F^4}{M^6} \, h(x)
\nn\\
&&
+\frac{1}{(4\pi)^4}
\Bigg\{
~~~~
2\left(3Tr\left(y_u^{\dagger}y_u \right)
-\frac{16}{3}g_3^2 - 3g_2^2 - \frac{13}{15}g_1^2\right) y_u^{\dagger}y_u
\nn \\
&&~~~~~~~~~~~~~~~~
+6 y_u^{\dagger}y_u y_u^{\dagger}y_u 
+2y_u^{\dagger}Y_u Y_u^\dagger y_u 
+2y_u^{\dagger}Y_d Y_d^\dagger y_u 
-2Y_u^\dagger y_uy_u^{\dagger}Y_u 
\nn \\
&&~~~~~~~~~~~~~~~~
+2y_u^{\dagger}Y_u Tr\left(3Y_u^\dagger y_u \right)
+2Y_u^\dagger y_u Tr\left(3y_u^{\dagger}Y_u\right)
~~~~\Bigg\}
\left|\frac{F}{M}\right|^2  \,,
\\
\delta\tilde m^2_{d_R}
&=&
-\frac{1}{(4\pi)^4}
2Y_d^\dagger y_u y_u^{\dagger}Y_d
\left|\frac{F}{M}\right|^2  \,.
\eeqa
where $x=\frac{F}{M^2}$ and~\cite{Evans:2011bea}
\beq
h(x) =- 3\frac{(2-x)\log(1-x) + (2+x)\log(1+x)}{x^4} 
\approx
1 + \frac{4x^2}{5} + {\cal O}(x^4)
\eeq
In addition, A-terms receive 1-loop contributions
\beqa
A^*_u
&=&
-\frac{1}{16\pi^2}\left[
\left(y_u y_u^{\dagger} \right)Y_u
+2Y_u\left( y_u^{\dagger}y_u \right) 
\right]\frac{F}{M} \,,
\\
A^*_d &=& 
-\frac{1}{16\pi^2}\left[
\left(y_u y_u^{\dagger} \right)Y_d
\right]\frac{F}{M} \,,
\\
A^*_l &=& 
0 \, .
\eeqa

Note that we have defined
\beqa
Y_u = Y^*_U\,,~~~ Y_d = Y^*_D\,,~~~Y_l = Y^*_L\,,~~~ y_u = y^*_U 
\eeqa
such that $Y_{u,~d,~l}$ coincide with the SM Yukawa matrices.


\providecommand{\href}[2]{#2}\begingroup\raggedright\endgroup

\end{document}